\documentclass[12pt]{article}
\usepackage{graphics,psfig,epsfig,multicol}
\usepackage{fancyhdr,deluxetable,supertab}
\usepackage{helvet}

\def\simlt{\mathrel{\hbox{\rlap{\hbox{\lower4pt\hbox{$\sim$}}}\hbox{$<$}}}}
\def\simgt{\mathrel{\hbox{\rlap{\hbox{\lower4pt\hbox{$\sim$}}}\hbox{$>$}}}}
\def\eps@scaling{.95}
\def\epsscale#1{\gdef\eps@scaling{#1}}
\def\plotone#1{\centering \leavevmode
\epsfxsize=\eps@scaling\textwidth \epsfbox{#1}}

\begin{document}
\thispagestyle{plain}
\begin{center}

{
\centering \leavevmode
\hspace{.02\columnwidth}
\epsfxsize=.46\columnwidth \epsfbox{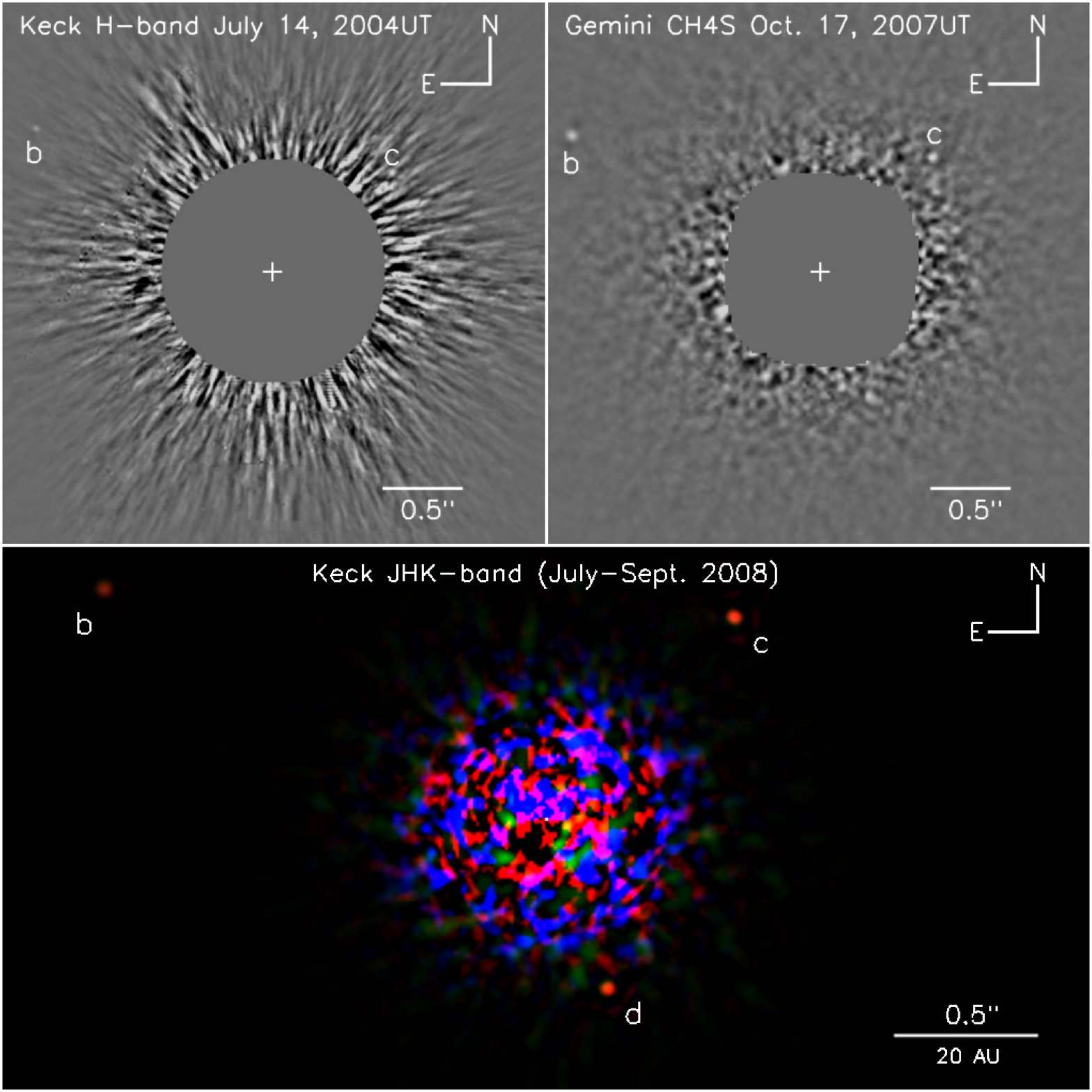}
\hspace{.02\columnwidth}
\epsfxsize=.46\columnwidth \epsfbox{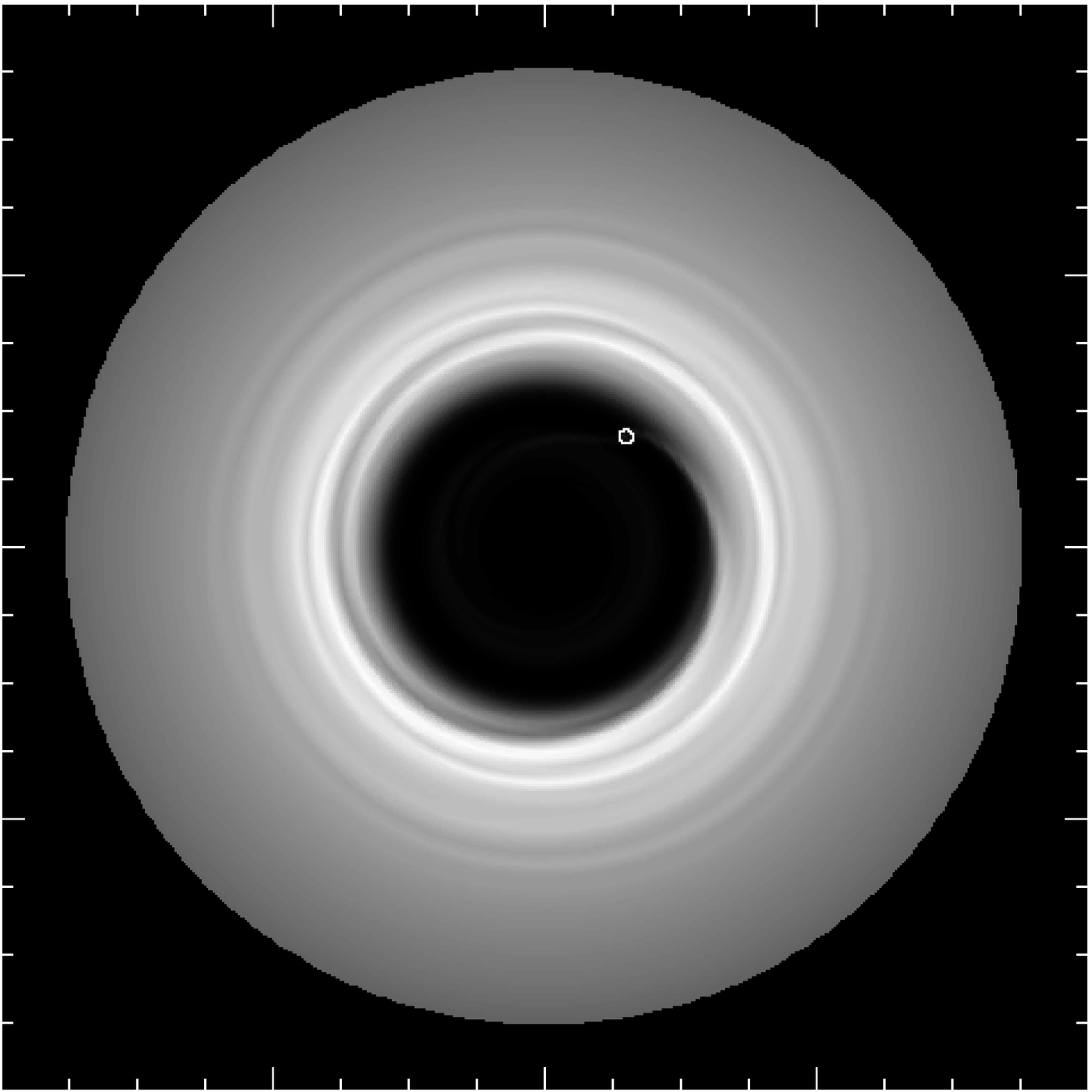}
}

\LARGE
{\bf The Formation and Architecture of\\
Young Planetary Systems}

\large
An Astro2010 Decadal Survey White Paper

\vspace{0.1in}

\large {\bf Adam Kraus} (Caltech), {\bf Kevin Covey} (Harvard/CfA),\\ 
{\bf Michael Liu} (Hawaii/IfA), {\bf Stanimir Metchev} (SUNY Stony 
Brook),\\ {\bf Russel White} (Georgia State), {\bf Lisa Prato} 
(Lowell),\\ {\bf Doug Lin} (UCSC), {\bf Mark Marley} (NASA Ames)

\normalsize

\vspace{0.1in}

{
\centering \leavevmode
\epsfxsize=.48\columnwidth \epsfbox{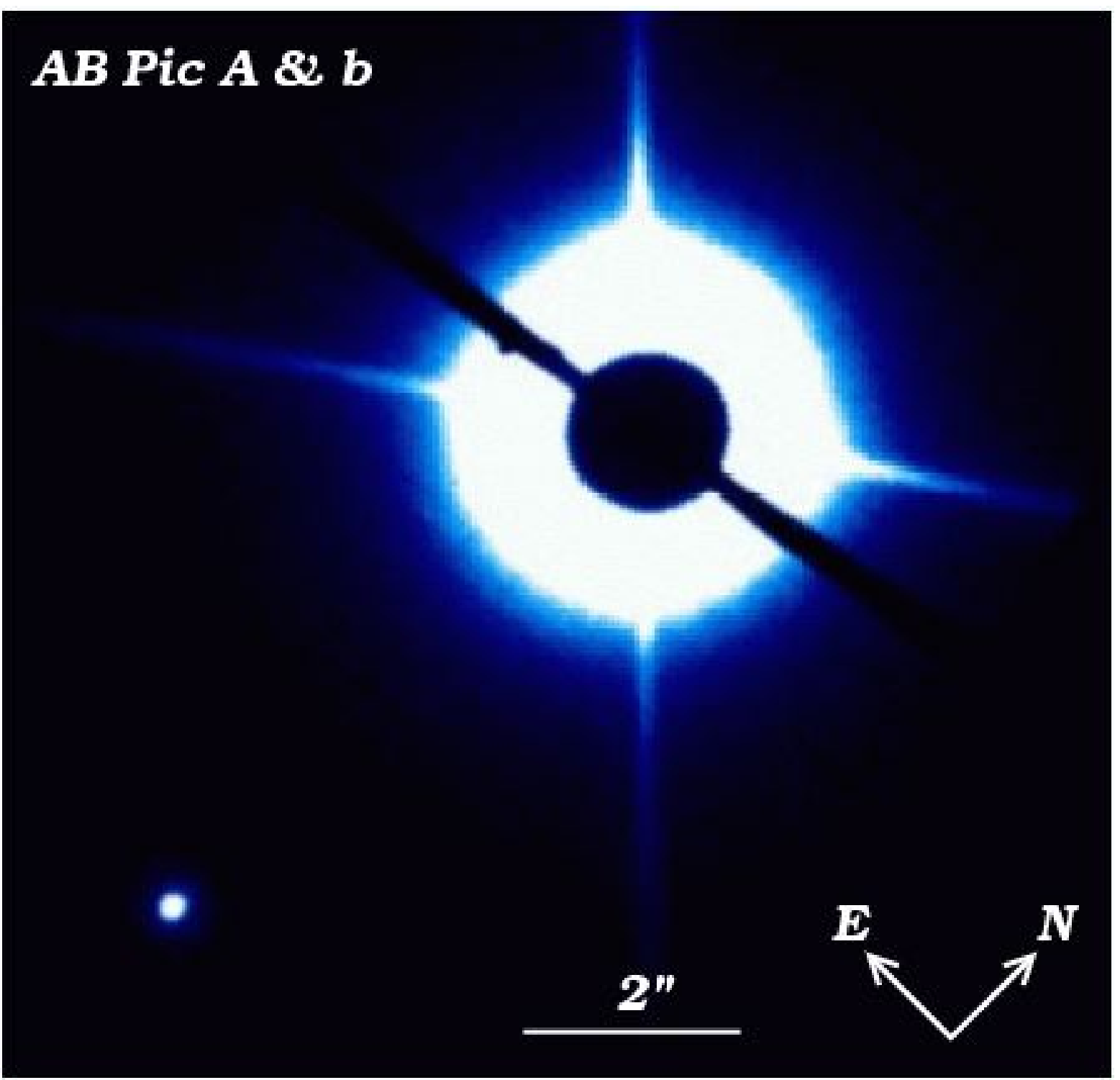}
\hspace{.02\columnwidth}
\epsfxsize=.46\columnwidth \epsfbox{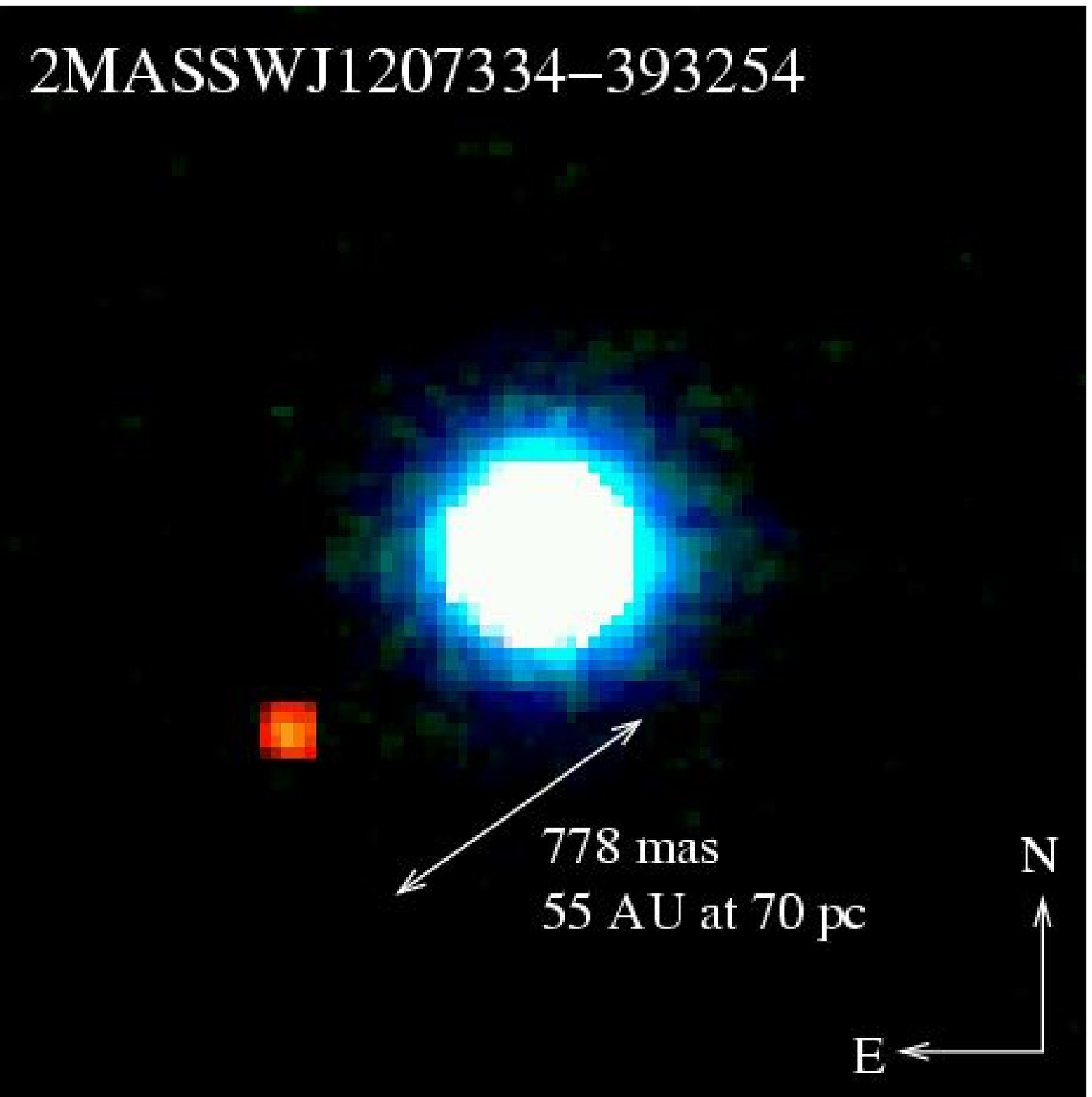}
}

 {
 \parbox[b]{0.48\columnwidth}{
 \footnotesize \it Clockwise from top left: The first directly-imaged 
extrasolar planetary system, HR 8799 bcd$^1$; a model of how a giant 
planet would clear a protoplanetary disk$^2$; the wide, apparently 
planetary-mass companions 2M1207b and AB Pic b$^{3,4}$.}
 \hspace{0.02\columnwidth}
 \parbox[b]{0.46\columnwidth}{
 \small
 Contact Information for Primary Author:\\
 A.L. Kraus, California Institute of Technology\\
 (626) 395-4095, alk@astro.caltech.edu\\}}

\end{center}
\clearpage

\noindent
\underline{\bf Abstract}
\indent

Newly-formed planetary systems with ages of $\simlt$10 Myr offer many 
unique insights into the formation, evolution, and fundamental 
properties of extrasolar planets. These planets have fallen beyond the 
limits of past surveys, but as we enter the next decade, we stand on 
the threshold of several crucial advances in instrumentation and 
observing techniques that will finally unveil this critical 
population. In this white paper, we consider several classes of 
planets (inner gas giants, outer gas giants, and ultrawide 
planetary-mass companions) and summarize the motivation for their 
study, the observational tests that will distinguish between competing 
theoretical models, and the infrastructure investments and policy 
choices that will best enable future discovery. We propose that there 
are two fundamental questions that must be addressed: 1) Do planets 
form via core accretion, gravitational instability, or a combination 
of both methods? 2) What do the atmospheres and interiors of young 
planets look like, and does the mass-luminosity relation of young 
planets more closely resemble the ``hot start'' or ``cold start'' 
models? To address these questions, we recommend investment in 
high-resolution NIR spectrographs (existing and new), support for 
innovative new techniques and pathfinder surveys for 
directly-imaged young exoplanets, and continued investment in 
visible-light adaptive optics to allow full characterization of 
wide ``planetary-mass'' companions for calibrating planet evolutionary 
models. In summary, {\bf testing newly proposed planet formation and 
evolutionary predictions will require the identification of a large 
population of young ($<$10 Myr) planets whose orbital, atmospheric, 
and structural properties can be studied.}

\vspace{0.05in}
\noindent
\underline{\bf Introduction}
\indent

The exciting discovery of extrasolar planets just over a dozen years 
ago has revitalized stellar and planetary science and generated a 
tremendous public interest in astronomy. Since then, an immense 
international effort has demonstrated that $\simgt$10-15\% of FGK-type 
stars harbor extrasolar giant planets$^5$. Surprisingly, however, the 
properties of many of these planets are radically different from the 
gas giant planets in our solar system.  Some orbit their host star at 
a small fraction of an AU in less than a week's time, while others 
have highly eccentric, binary star-like orbits$^{6,7}$.  Planet-like 
companions (if evolutionary models are correct), have even been 
directly imaged at separations of more than 100 AU from their host 
star $^{3,4,8}$.  These unexpected properties forced sweeping changes 
in the standard picture of how disk material assembles into 
planets$^9$. Unfortunately, testing these new theories is extremely 
difficult because of the lack of direct observational constraints.

Currently, the two competing paradigms for forming extrasolar giant 
planets (EGPs) are the core accretion and gravitational instability 
models$^{10,11}$. Core accretion provides a natural explanation for 
the enhanced EGP frequency around metal-rich stars$^{12,13}$ and the 
massive solid cores of many EGPs$^{14,15,16}$. Also, core accretion 
should proceed most rapidly at the snow line, with subsequent 
migration of the resulting gas giants to smaller radii, so the model 
naturally explains the gas giant population at $<$5 AU discovered by 
field RV surveys. However, the core accretion timescale is 
unrealistically long at radii of $>$10 AU$^{17}$, so it can not 
explain wide exoplanets like Fomalhaut b$^8$ and HR 8799 bcd$^1$, much 
less the ultrawide planetary-mass companions like 2M1207 b$^3$. The 
rapid collapse and growth of gravitational instabilities in a 
protostellar or protoplanetary disk provides a more feasible 
explanation for the formation of wide systems, so it appears that both 
processes contribute to the overall exoplanet population. Surveys for 
young exoplanets are critical for distinguishing the relative 
importance of each process.

The atmospheres and interior structure of young exoplanets are also 
completely unconstrained by current observations. The 
mass-luminosity-age relation is very sensitive to the formation 
scenario (the ``hot start'' versus ``cold start'' 
models)$^{18,19,20}$. The predicted luminosities vary by as much as 
2-3 orders of magnitude, so the first empirically measured fundamental 
properties of young exoplanets (i.e. temperatures and luminosities) 
will provide an unambiguous endorsement of one set of models. The 
predicted yield from future direct imaging surveys, and therefore the 
relative merits of imaging and astrometric missions, depends 
critically on the assumed mass-luminosity relation. The first handful 
of ground-based detections with existing technology will allow us to 
finally calibrate young EGP models, determining the missions that 
should be supported for 2020 and beyond.

Both of these open questions must be addressed at the age range when 
planets form, before evolution obscures the signatures of their 
formation mechanism (via migration and planet-planet interactions) and 
primordial interior structure (via radiation of the primordial energy 
from assembly). This age range is set by the disk dissipation 
timescale ($\simlt$10 Myr)$^{21}$ and the timescale for the 
``hot-start'' and ``cold-start'' models to converge ($\sim$10 Myr at 2 
$M_{Jup}$)$^{19}$, encompassing most of the star-forming regions in 
the solar neighborhood. One notable feature is the large distance to 
these populations; aside from a few sparse moving groups, all stars 
younger than $\sim$10 Myr are located at distances of $\simgt$120 pc, 
which strongly limits the choice of observing strategies.

Ground-breaking results from Spitzer have revolutionized our 
understanding of protoplanetary disk formation, and the rapid pace for 
discovery of nearby exoplanets has transformed our understanding of 
old planetary systems. However, the difficulty of identifying young 
planets has left us few clues on the early evolutionary processes that 
transform protoplanetary disks into architecturally mature systems; 
{\it as we begin the next decade, the field has yet to identify even 
one young exoplanet}. In this white paper, we describe the science 
drivers and observational goals that should guide young exoplanet 
science in the coming decade. The critical questions and observations 
fall into three regimes that will be probed via different techniques: 
the inner solar system (via radial velocity surveys), the outer solar 
system (via extreme AO imaging and interferometry), and ultrawide 
substellar companions (via deep coronagraphic imaging). Our policy 
recommendations (bold face at section ends) are aimed at ground-based 
surveys; recommendations for space surveys can be found in papers by 
Beichman (SIM) and Sivaramakrishnan (JWST).

{
\centering \leavevmode
 \epsfxsize=.54\columnwidth \epsfbox{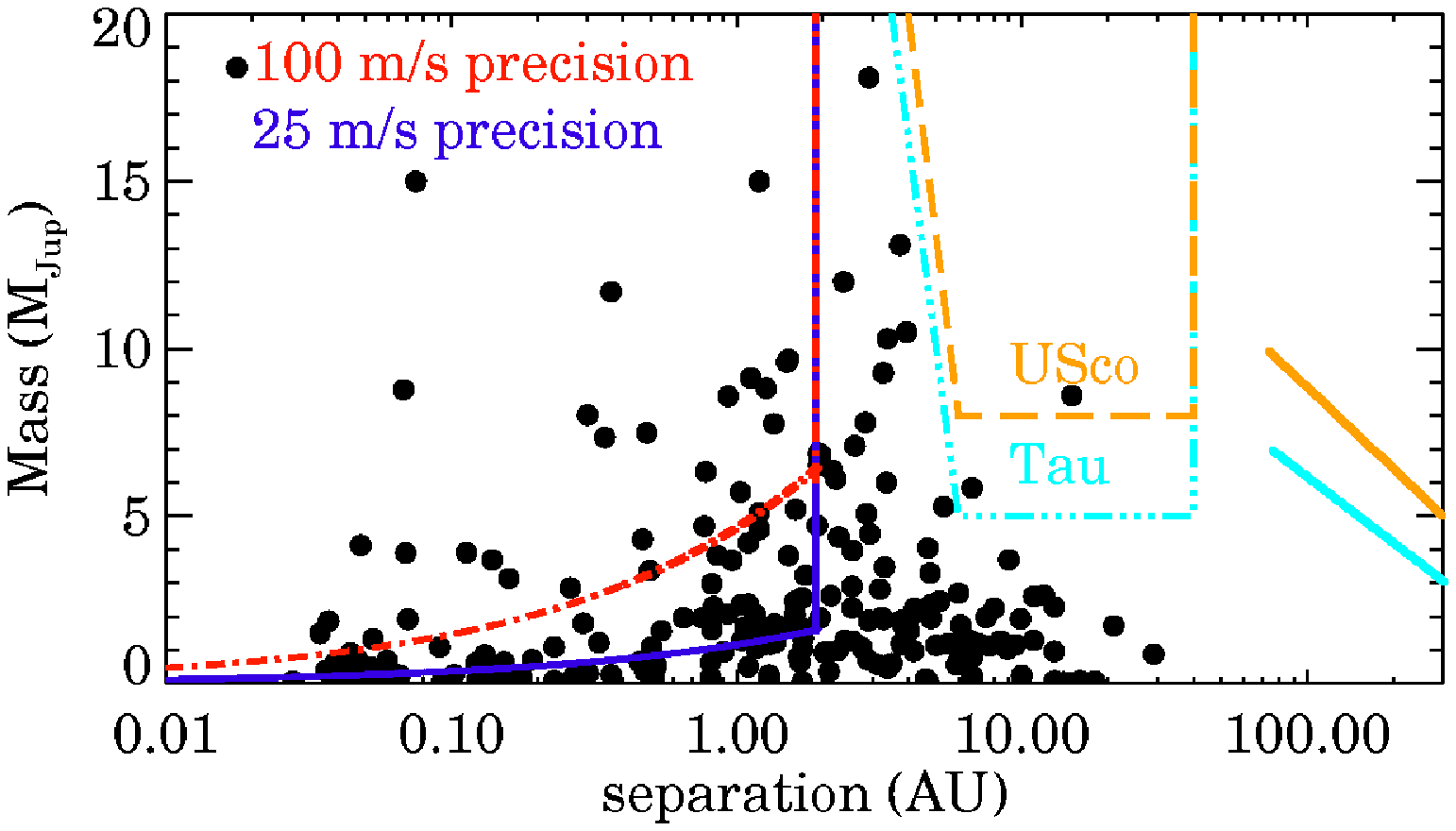}
 \parbox[b]{0.44\columnwidth}{
 \footnotesize {\bf Figure 1.} Projected limits for the survey methods 
we describe. For RV surveys, we show the limits corresponding to 
measurement precisions of 100 m/s (red) and 25 m/s (blue). For 
aperture masking, we show the limits in Taurus (1-2 Myr; dashed cyan) 
and Upper Sco (5 Myr; dashed orange) given the achieved contrast limit 
($\Delta$$L=6-7$ at $\Lambda$$/D$) and the models of Baraffe et al. 
(2002). For wide PMCs we show similar limits (solid lines) for 
obtaining a spectrum with $S/N\sim$10 and $R\sim$1000 in 6 hours on a 
10m telescope with visible-light AO.}}

\clearpage

\footnotesize
\noindent
{\bf Table 1.} {\it Open Problems and Solutions for the Next Decade}\\
\begin{tabular}{|l|l|l|}
\hline
\it Observable Test&\it Theoretical Constraint&{\it Observation}\\
\hline
How quickly do planets form?&Core Accretion or Grav. Instability?&RV, Imaging\\
\hline
Planets at wide separations or the snow line?&Core Accretion of 
Grav. Instability?&Imaging\\
\hline
How bright are the most massive planets?&``Hot Start'' or ``Cold Start'' 
Models?&Imaging\\
\hline
When do Hot Jupiters appear?&Constraining Migration Timescales&RV\\
\hline
Do planets open most of the gaps in disks?&How ubiquitous are planets?&Imaging\\
\hline
What are the properties of wide ``planets''?&``Hot Start'' or ``Cold 
Start'' 
Models?&Visible-Light AO\\
\hline

\end{tabular}
\normalsize

\vspace{0.15in}

\noindent 
\underline{\bf Revealing Hot Jupiters Through High-Precision Infrared Spectroscopy}
\indent

\textit{Determining the frequency of young hot Jupiters, as well as their 
physical and orbital properties, will provide new, empirical constraints 
for models of planet formation and migration.} Even accounting for errors 
in age estimates for young stars$^{22}$, it will be hard to reconcile a 
dearth of planets around stars younger than 3 Myrs with the gravitational 
instability model, which predicts planet formation timescales much 
shorter than 1 Myr$^{11}$.  The distribution of orbital separations and 
eccentricities, and their evolution with age, will provide a strong test 
for models of planetary migration; similarly, if the formation of gas 
giant planets is indeed the dominant mechanism driving circumstellar disk 
dispersal, then we expect an elevated planet frequency for stars whose 
disks have large inner holes or gaps. Finally, one of the most compelling 
reasons to search for young hot Jupiters is that $\sim$1/10 will transit 
their host star. The size and density of a planet are strictly 
constrained by its transit depth and shape, and its atmosphere can be 
studied through transmission spectroscopy$^{23}$. Even a handful of these 
systems would yield unprecedented constraints on the composition and 
structure of planets immediately after formation.

Building a significant sample of young hot Jupiters, as required to 
define the frequency and timescale of gas giant formation, will only be 
possible through precise radial velocity (RV) monitoring in the 
near--infrared. Most young stars are too distant (d$\ge$120 pc) to detect 
planets within $\sim$1 AU via direct imaging or astrometry, and 
photometric variability due to accretion and flares make transit 
detections virtually impossible.  Most perniciously, starspots distort 
optical spectral lines as they rotate across the surface of young stars, 
mimicking RV signals due to planets$^{24}$. The young, actively accreting 
classical T Tauri star TW Hydrae provides a cautionary tale: RV 
periodicity was detected in the optical and, after careful scrutiny, 
attributed to a planet$^{25}$. Follow-up observations with CRIRES, a high 
resolution (R$\sim$100,000) near-infrared (NIR) spectrograph on the VLT, 
found that TW Hydrae's RV signature is strongly wavelength dependent and 
disappears entirely in the H band (Figure 1)$^{26}$. These observations 
corroborate predictions that RV anomalies are minimized in the NIR due to 
lowered contrast between Rayleigh-Jeans emission from starspots and the 
surrounding photosphere; Prato et al. recently confirmed this 
effect$^{27}$.

While NIR Doppler observations indicate that TW Hydrae does not host a 
gas giant, the star's stability at 35 m/s precision demonstrates that 
this technique will detect young gas giants elsewhere. Accumulating 
even a moderately sized sample of young planets, however, will require 
a significant investment of observational resources; if young stars 
host planets with the same frequency as nearby solar analogs this will 
require monitoring hundreds of young stars. Although the global 
astronomical community currently lacks the capacity for a survey of 
this scale, pioneering efforts with available (but outdated) US 
facilities demonstrate the feasibility. RV surveys have obtained 
precisions of 100 m/s

\clearpage

{
\centering \leavevmode
 \epsfxsize=.51\columnwidth \epsfbox{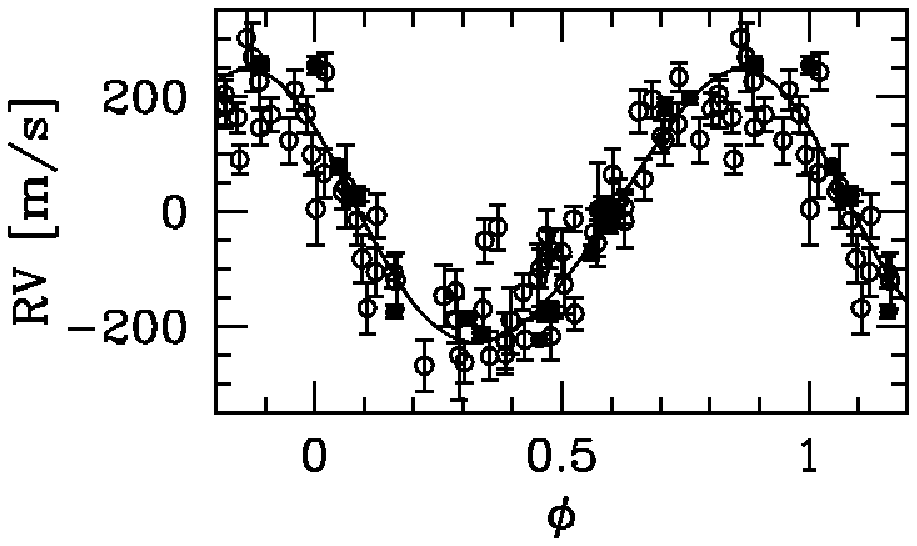}
 \epsfxsize=.44\columnwidth \epsfbox{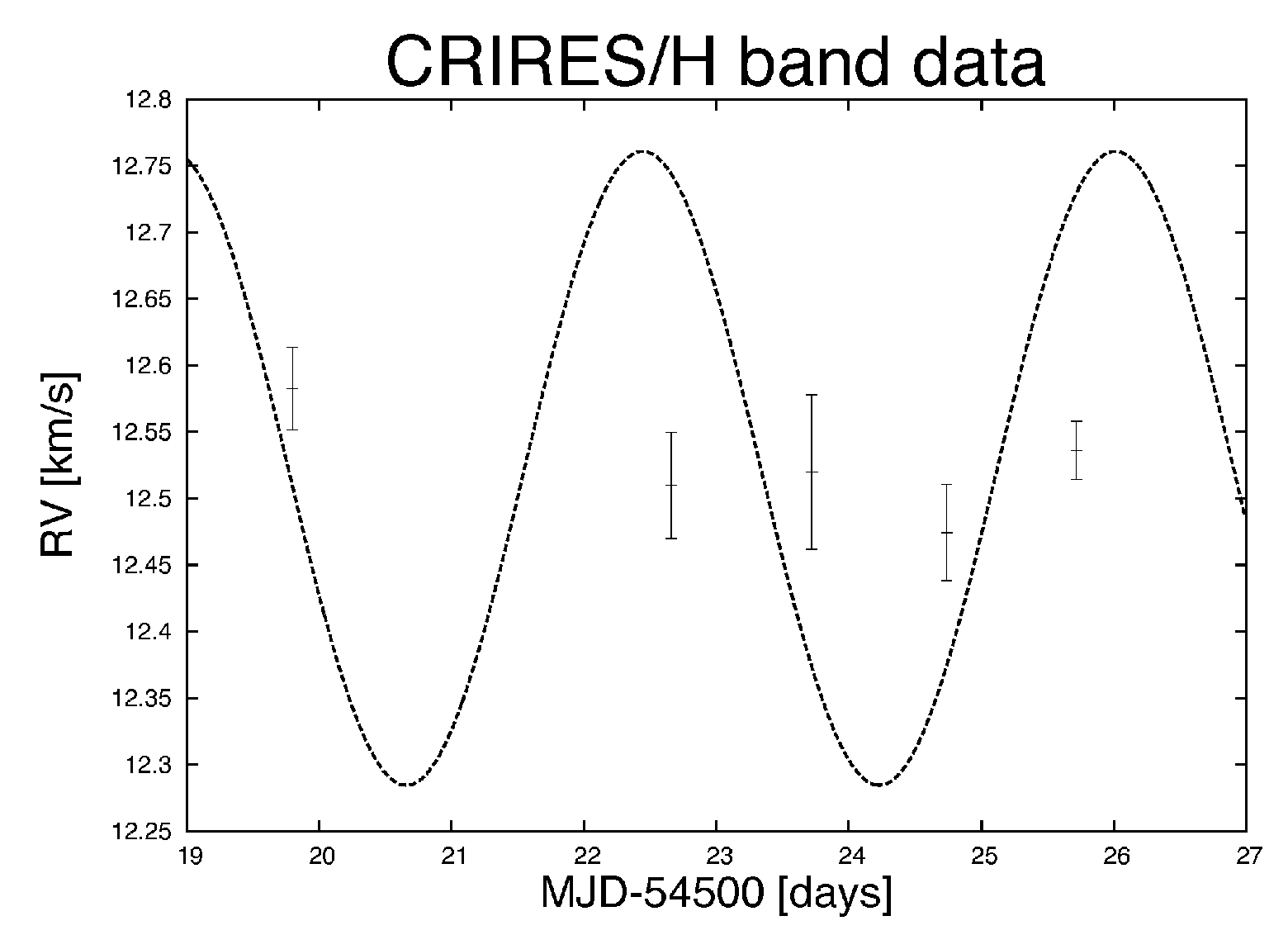} 
}

\vspace{-0.05in}
\noindent \footnotesize {\bf Figure 2.} Left: Phase-folded RV 
measurements of TW Hya obtained with the optical spectrographs HARPS, 
CORALIE, and FEROS. The best Keplerian fit has a semi-amplitude of 238 
m/s, and was initially interpreted as reflex motion due to a 
planet$^{25}$. Right: Infrared RV measurements of TW Hya obtained with 
CRIRES show that its RV is constant to within 35 m/s and suggest the 
variations seen at optical wavelength are due to star spots.$^{26}$ 
(Figures from Hu\'elamo et al. 2008). \normalsize

\vspace{0.13in}

\noindent with CSHELL at IRTF$^{27}$ and 50 m/s with 
NIRSPEC at Keck$^{28}$; next generation instruments should yield even 
better results$^{29,30,31}$. Nevertheless, the demonstrated stability 
of CRIRES on the VLT and the upcoming commissioning of NAHUAL on the 
Gran Telescopio de Canarias$^{32}$ present a challenge for continued 
US competitiveness. {\bf Ensuring that U.S. investigators lead this 
area over the coming decade will require significant community access 
to existing high-dispersion NIR spectrographs and development of new 
instruments to expand the U.S. capacity for measuring precise RVs in 
the NIR.}

\vspace{0.1in}
\noindent 
\underline{\bf Unveiling Outer Planetary Systems with Innovative Direct-Imaging Techniques}
\indent

{\it The frequency and properties of outer extrasolar giant planets 
(EGPs) at $a\sim$5--40 AU will provide crucial constraints on the 
process and ubiquity of planet formation.} Direct imaging surveys 
present the only realistic prospect for studying these long-period 
planets since RV and astrometric surveys would require decades-long 
monitoring campaigns. These surveys are best conducted for very young 
systems since young EGPs are more luminous than their older brethren, 
significantly reducing the contrast between stars and planets, though 
the distance to nearby star-forming regions ($d\ge$120 pc) imposes a 
corresponding resolution penalty. Planetary systems undergo 
significant dynamical evolution after birth, so it is also important 
to study the most architecturally pristine systems.

Planet formation models now form Jupiter and Saturn in situ, via core 
accretion in $\sim$3 Myr, which is also the disk dissipation 
timescale$^{21,33}$. These competing timescales make the outer EGP 
frequency a sensitive test of formation models, as significantly 
changing either would make outer EGPs scarce or ubiquitous. The planet 
formation timescale also distinguishes between formation models; a 
paucity of EGPs at $<$2-3 Myr would argue that planets form slowly via 
core accretion, whereas the existence of planets around the youngest 
stars would require a prompt process like gravitational instability. 
Finally, core accretion is fastest at the snow line$^{34}$, indicating 
that many EGPs should form at $\sim$3-5 AU and migrate inward. A 
broader orbital distribution would argue that some EGPs form via 
gravitational instability, though this test is only significant at 
young ages; interactions with planetesimals or other planets can 
scatter EGPs outward within $\sim$50-100 Myr$^{35}$. The 
distribution of EGPs also will reveal the gaps within which terrestrial 
planets could form.

Ongoing planet/disk interactions will also provide new insight into 
the planet formation process. Many nearby young stars show firm 
observational evidence of gaps or inner holes with radii of 
10-40 AU (see cover figure); these ``transitional disk'' systems are 
identified from SED modeling$^{36,37,38}$ or spatially resolved 
submm/mm imaging$^{39}$. If these gaps are signposts of ongoing planet 
formation, then they show exactly where to search for the youngest, 
most luminous exoplanets. Furthermore, if most gaps host planets, it 
would indicate that a large fraction of disks are cleared by planet 
formation; empty gaps might indicate that photoevaporation clears most 
disks, forestalling EGP formation.

The current generation of planet-search instruments, including GPI and 
SPHERE, are optimized to deliver high contrast ($\sim$10$^4$-10$^6$) 
at wide separations ($>$0.5'') to search for planets around nearby 
stars. Young planets are much brighter, but because their host stars 
are more distant, most planets will fall inside the instruments' 
coronagraph radius ($\sim$200 mas or $\sim$30 AU). In the long term, 
extreme AO systems on ELTs could identify Jupiter analogues ($a\sim$5 
AU; $M\sim$1 $M_{Jup}$) around young stars. However, this decade will 
be dominated by existing telescopes that use new instruments or 
techniques to search for massive Jupiter analogs ($a\sim$5-10 AU; 
$M\sim$5-10 $M_{Jup}$).

One of these promising techniques is aperture-mask 
interferometry$^{40,41,42}$ (Figure 3), which achieves superior 
contrast limits over imaging at small separations ($\sim$10$^2$-10$^3$ 
at $\Lambda$$/D$) by resampling a single telescope aperture into a 
sparse interferometric array. This technique can achieve detection 
limits of $\sim$7-10 $M_{Jup}$ at ages of 5 Myr and $\sim$5 $M_{Jup}$ 
at 1 Myr. There are also ongoing plans for a masking survey with JWST 
that will achieve contrasts of 10$^4$--10$^5$ (see paper by 
Sivaramakrishnan). On a $\sim$3-5 year timeframe, extreme AO systems 
on existing telescopes will also be commissioned, including PALM3K at 
Palomar and NGAO on Keck. The superior AO performance with respect to 
current imaging techniques could surpass aperture-mask interferometry. 
The continued availability of telescope time for these surveys will 
also be critical; most current initiatives are using private 
facilities like Keck and Palomar, but to remain competitive with 
ongoing European programs at the VLT, we must open the field to the 
entire US community with either increased access to the private 
observatories or improved flexibility in bringing outside technologies 
to Gemini. {\bf The mass-luminosity relation of young exoplanets and 
the population statistics of outer gas giants are undetermined, so 
pathfinder surveys that exploit new instruments and techniques at 
existing observatories will be critical in guiding next-generation 
missions from space and with the TMT/GMT.}

\vspace{0.1in}
{
\centering \leavevmode
 \epsfxsize=.55\columnwidth \epsfbox{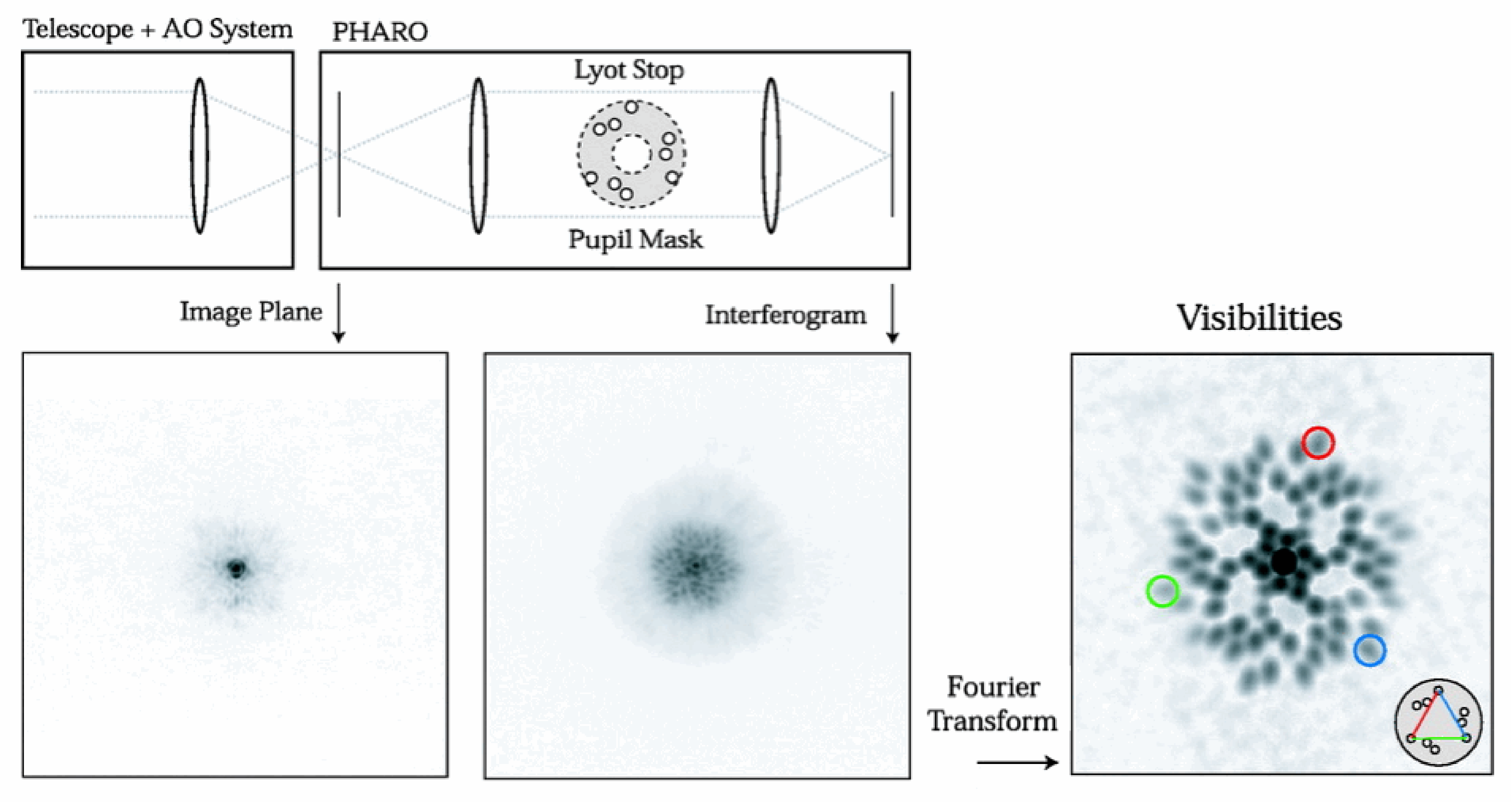}
 \hspace{0.01\columnwidth}
 \parbox[b]{0.44\columnwidth}{
 \footnotesize {\bf Figure 3.} Schematic of aperture-mask interferometry, 
one experimental high-resolution imaging technique implemented at Keck and 
Palomar. A mask transforms the single aperture into an interferometric 
array, yielding an interferogram. A Fourier transform recovers the 
visibilities. This technique represents only one opportunity for 
discovering exoplanets; {\bf openness to experimental instruments and 
techniques is one of the strongest advantages of our private 
observatories and should be extended to our national system.}}}

\clearpage

{
\centering \leavevmode
 \epsfxsize=.52\columnwidth \epsfbox{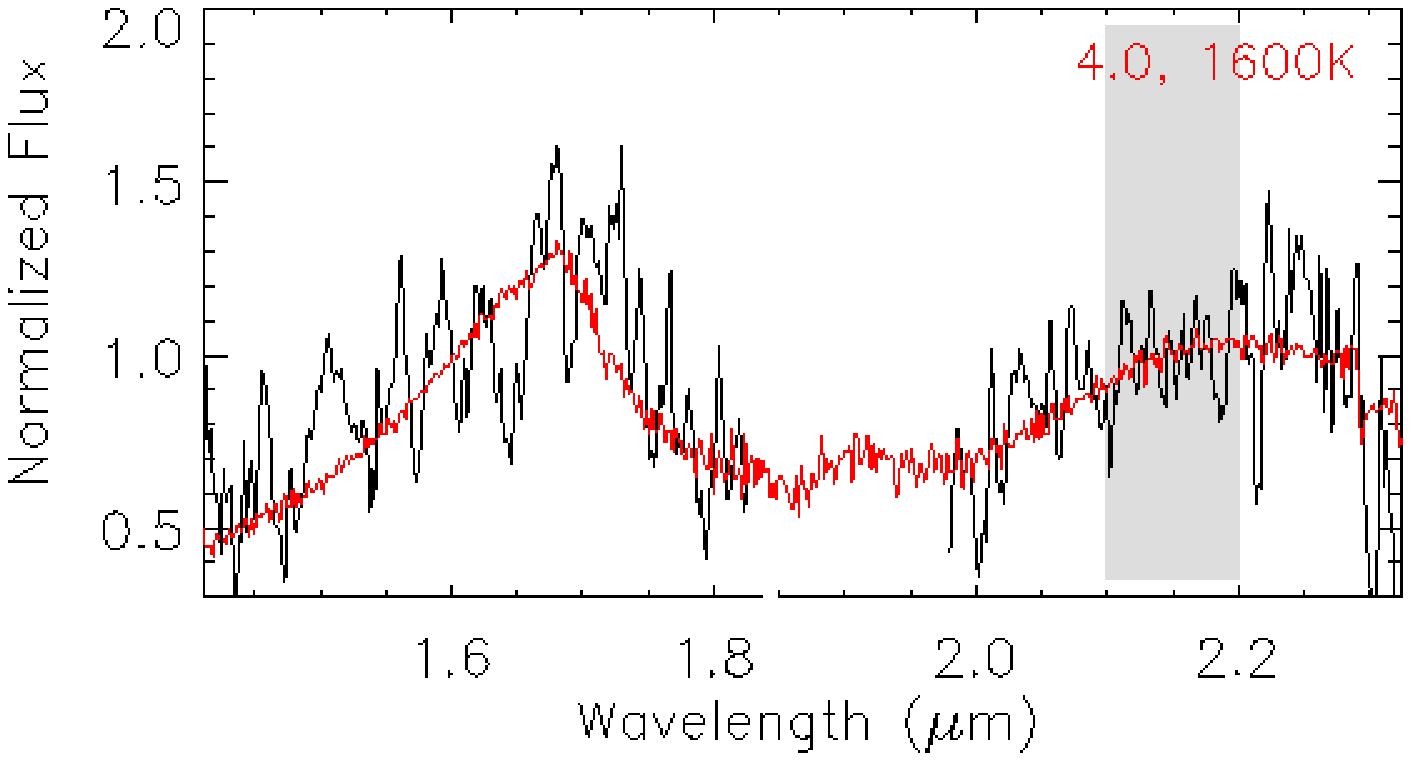}
 \hspace{0.07\columnwidth}
 \epsfxsize=.40\columnwidth \epsfbox{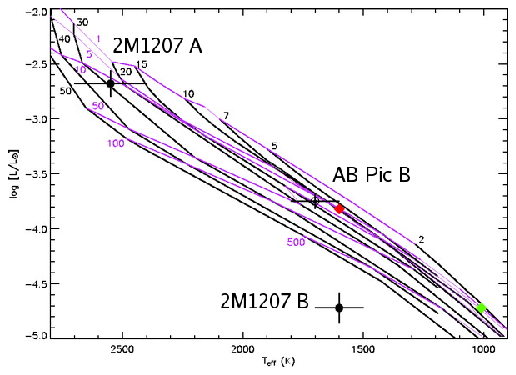}
}

\vspace{-0.25in}
\noindent \footnotesize {\bf Figure 4.} Left: NIR spectrum of 2M1207b 
(black), an apparently planetary-mass companion with an age of $\sim$8 
Myr and a spectral type of $\sim$L6; models predict a corresponding 
mass of $\sim$5-8 $M_{Jup}$. The red line shows a SETTL model spectrum 
with $\log g$$=4.0$ and $T_{eff}$$=1600$ K.$^{46}$ Right: The 
positions of 2M1207b and AB Pic b on an HR diagram, plus the 
isochronal (black) and isomass (purple) lines of the Lyon models 
(Baraffe et al 2002). The position of AB Pic b agrees very well with 
its age, but 2M1207b is significantly underluminous. {\bf Placing a 
planetary-mass object on an HR diagram should be regarded as an 
outstanding success for our field, but significant uncertainties in 
their properties and provenance must be addressed in the coming 
decade.$^{46}$} (Figures and results from Mohanty et al. 2007). 
\normalsize

\vspace{0.05in}
\noindent
\underline{\bf Calibrating Young Exoplanet Models with Wide Planetary-Mass Companions}
\indent

Over the past five years, direct imaging surveys for extrasolar 
planets have discovered a small but significant number of 
planetary-mass companions (PMCs) at $>$50 AU separations from their 
primaries (see cover page). The prototypical system, 2M1207-3933, 
consists of a 4 $M_{Jup}$ companion located $\sim$50 AU away from a 10 
Myr old brown dwarf$^{3}$. Since its discovery, $\sim$5 other PMCs 
have also been reported, most of which orbit much higher-mass 
primaries ($\sim$0.5--1.5 $M_{\odot}$). PMCs pose a significant 
challenge to existing models of planet formation. The core accretion 
timescale ($>>$100 Myr at 100 AU$^{17}$) is far longer than the disk 
dissipation timescale ($\simlt$3-5 Myr$^{21}$). Gravitational 
instability could form PMCs$^{11}$, but only for disks that dwarf the 
most massive systems currently observed ($\sim$0.05 
$M_{\odot}$$^{43}$). Binary formation also is unable to explain PMCs, 
as theoretical simulations are unable to produce extremely 
unequal-mass companions$^{44}$ and PMCs are too common to represent 
the extreme tail of the observed binary mass function$^{42,45}$.

Wide PMCs are far easier to study than their analogs in normal 
planetary systems, so their atmospheres and interiors provide an 
empirical baseline for models of young gas giant planets. The 
luminosities of young exoplanets are currently uncertain by as much as 
$\sim$2 orders of magnitude$^{18,19,20}$, so it is critical to 
determine whether these companions are genuinely 5-15 $M_{Jup}$ (as is 
predicted by mass-luminosity relations). This issue is complicated by 
PMCs' uncertain origin; if they form via binary fragmentation or via 
gravitational instability in a massive protostellar disk, then they 
might not have the same interior structure and evolutionary history as 
conventional exoplanets.

Past studies of PMCs were limited to NIR photometry and spectroscopy 
(Figure 4) due to their extreme faintness and the current limits of 
AO. NIR techniques are sufficient for PMC discovery, but new results 
for young brown dwarfs show that NIR SEDs are severely impacted by 
condensate cloud levels, leaving NIR observations weakly diagnostic of 
PMCs' physical properties$^{47,48}$. Optical fluxes and spectra are 
much less affected, yielding accurate measurements of temperature, 
metallicity, and surface gravity, but they are difficult to acquire. 
Current AO systems do not operate shortward of 1$\mu$m, and the small 
aperture of HST is insufficient for optical spectroscopy. It is also 
important to study indicators of their formation process such as 
accretion (from H$\alpha$ emission) or disk evolution (from atypical 
SEDs due to reflected starlight, as for Fomalhaut b)$^{8}$.

Coupled with these empirical tests is the need for theoretical 
advances in atmospheric models themselves. Beyond the complex 
processes of condensate cloud formation and chemistry in dynamic 
atmospheres, models also require improvements in chemical abundances 
and opacities. Currently, there are large uncertainties in the 
absolute abundances of CNO in our reference standard, the 
Sun$^{49,50}$. CNO-bearing molecules are a dominant source of opacity 
in planetary atmospheres, so abundance uncertainties translate into 
systematic uncertainties in atmosphere models. Models also suffer from 
incomplete warm opacity line lists for key molecules; line lists at 
wavelengths $<$1.6 $\mu$m are incomplete for CH$_4$ and nonexistent 
for NH$_3$. These molecules produce strong absorption bands that leads 
to substantial flux redistribution, propagating spectral modeling 
errors to other wavelengths. Calculating the very large number of 
transitions for these molecules is computationally intensive, while 
the supporting laboratory studies remain challenging.

Existing telescopes and instruments are sufficient for the continued 
discovery of very wide planetary-mass companions, but we lack the 
capabilities needed to characterize their fundamental properties and 
formation history. Visible-light AO systems on large-aperture telescopes, 
capable of both imaging and spectroscopy, will be crucial in extending our 
studies into the optical regime; initiatives like PALM3K at Palomar and 
NGAO at Keck will lead the field.  Visible-light AO will also be required 
to study detailed accretion processes and reflected light from 
circumplanetary disks; further advances in visible-light AO will 
eventually even allow the direct study of reflected optical light from 
young planets at smaller radii. {\bf Resolved planetary-mass companions 
will provide templates for the atmospheric properties of all young 
exoplanets, so we must support the observational advances in visible-light 
AO and computational advances in atmospheric physics that are needed to 
characterize their formation, atmospheres, and interiors.}

\vspace{-0.15in}
\begin{multicols}{2}
\noindent
\scriptsize
\underline{\bf References}

\noindent
1. Marois, C. et al. 2009, Science, ref

\noindent
2. Quillen, A. et al. 2004, ApJ, 612, 137

\noindent
3. Chauvin, G. et al. 2004, A\&A, 425, 29  

\noindent
4. Chauvin, G. et al. 2005, A\&A, 438, 29   

\noindent
5. Butler, R.P. et al. 2006, ApJ, 646, 505 

\noindent
6. Fischer, D. et al. 2008, ApJ, 675, 790

\noindent
7. Jones, H. et al. 2006, MNRAS, 369, 249

\noindent
8. Kalas, P. et al. 2008, Science, 322, 1345

\noindent
9. Santos, N. et al. 2005, Science, 310, 251

\noindent
10. Ida, S. \& Lin, D. 2004, ApJ, 616, 567

\noindent
11. Boss, A. 2001, ApJ, 563, 367

\noindent
12. Gonzalez, G. 1997, MNRAS, 285, 403

\noindent
13. Fischer, D. \& Valenti, J. 2005, ApJ, 622, 1102

\noindent
14. Saumon, D. \& Guillot, T. 2004, ApJ, 609, 1170

\noindent
15. Sato, B. et al. 2005, ApJ, 633, 465

\noindent
16. Guillot, T. et al. 2006, A\&A, 453, 21

\noindent
17. Pollack, J. et al. 1996, Icarus, 124, 62

\noindent
18. Baraffe, I. et al. 2003, A\&A, 402, 701

\noindent
19. Marley, M. et al. 2007, ApJ, 655, 541

\noindent
20. Fortney, J. et al. 2008, ApJ, 683, 1104

\noindent
21. Haisch, K. et al. 2001, ApJ, 553, 153

\noindent
22. Hillenbrand, L. \& White, R. 2004, ApJ, 604, 741

\noindent
23. Charbonneau, D. et al. 2002, ApJ, 568, 377

\noindent
24. Huerta, M. et al. 2008, ApJ, 678, 472

\noindent
25. Setiawan, J. et al. 2008, Nature, 451, 38

\noindent
26. Hu\'elamo, N. et al. 2008, A\&A, 489, 9

\noindent
27. Prato, L. et al. 2008, ApJ, 687, 103

\noindent
28. Bailey, J. et al. 2009, in prep

\noindent
29. Edelstein, J. et al. 2007, SPIE, 6693, 26

\noindent
30. Jones, H. et al. 2008, SPIE, 7014, 31

\noindent
31. Ramsey, L. et al. 2008, PASP, 120, 887

\noindent
32. Mart\'in, E. et al. 2005, AN, 326, 1026

\noindent
33. Dodson-Robinson, S. et al. 2008, ApJ, 688, 99

\noindent
34. Ida, S. \& Lin, D. 2008, ApJ, 685, 584

\noindent
35. Levison, H. et al. in "Protostars \& Planets V", 669

\noindent
36. Calvet, N. et al. 2005, ApJ, 630, 185

\noindent
37. Espaillat, C. et al. 2007, ApJ, 670, 135

\noindent
38. Brown, J. et al. 2007, ApJ, 664, 107

\noindent
39. Brown, J. et al. 2008, ApJ, 675, 109

\noindent
40. Tuthill, P. 2000, PASP, 112, 555   

\noindent
41. Lloyd, J. et al. 2006, ApJ, 650, 131

\noindent
42. Kraus, A. et al. 2008, ApJ, 679, 762

\noindent
43. Andrews, S. \& Williams, J. 2005, ApJ, 631, 1134

\noindent
44. Bate, M. 2009, MNRAS, 392, 590

\noindent
45. Metchev, S. \& Hillenbrand, L., ApJS, in press

\noindent
46. Mohanty, S. et al. 2007, ApJ, 657, 1064

\noindent
47. Cruz, K. et al. 2009, AJ, 137, 3345  

\noindent
48. Herczeg, G. et al. 2009, ApJ, in press 

\noindent
49. Asplund, M. et al. 2005, ASPC, 336, 25

\noindent
50. Lodders, K. et al. 2009, A\&A, in press

\end{multicols}

\end{document}